\documentclass[runningheads]{llncs}
\usepackage[T1]{fontenc}
\usepackage{graphicx}
\usepackage{listings}
\usepackage{amsmath}
\usepackage{amssymb}
\usepackage{graphicx}
\usepackage{bm}
\usepackage{color}
\usepackage{authblk}
\usepackage{booktabs}
\usepackage{multirow}
\usepackage{float}
\usepackage{tikz}
\usepackage{ctable}
\usepackage[colorlinks, citecolor=green]{hyperref}
\newcommand\tstrut{\rule{0pt}{2.4ex}}

\usepackage{color, colortbl}
\definecolor{Gray}{gray}{0.9}
\usetikzlibrary{arrows.meta, arrows}
\usepackage{diagbox}
\usepackage{breakcites}
\usepackage{orcidlink}
\usepackage{setspace}
\usepackage{tabularx}
\usepackage{subcaption}

\begin{document}

\title{Unsupervised whole-heart function assessment}
\author{
Yundi Zhang\inst{1,2}\orcidlink{0009-0008-7725-6369} \and
Daniel~Rueckert\inst{1,2,3}\orcidlink{0000-0002-5683-5889} \and
Jiazhen Pan\inst{1,2}\orcidlink{0000-0002-6305-8117}
}
\authorrunning{Y. Zhang et al.}

\institute{School of Computation, Information and Technology, Technical University of Munich, Germany
\and School of Medicine, Klinikum Rechts der Isar, Technical University of Munich, Germany
\and Department of Computing, Imperial College London, UK
\\
\email{\{yundi.zhang,jiazhen.pan\}@tum.de}
}

\maketitle              

\section{Synopsis}
\noindent \underline{\textbf{Motivation:}} CMR is the golden standard for cardiac diagnosis, and medical data annotation is time-consuming. Thus, screening techniques from unlabeled data can help streamline the cardiac diagnosis process.

\noindent \underline{\textbf{Goal(s):}} This work aims to enable cardiac function assessment from unlabeled cardiac MR images using an unsupervised approach with masked image modeling.

\noindent \underline{\textbf{Approach:}} Our model creates a robust latent space by reconstructing sparse 2D+T planes (SAX, 2CH, 3CH, and 4CH views) with 70\% masking, which can be further disentangled into distinct cardiac temporal states.

\noindent \underline{\textbf{Results:}} t-SNE visualization and kNN clustering analysis confirm the association between latent space and cardiac phenotypes, highlighting strong temporal feature extraction.

\noindent \underline{\textbf{Impact:}} This method offers a scalable approach for cardiac screening by creating a latent space as well as distinct time-segment embeddings, enabling diverse preliminary analysis of cardiac function and potentially advancing research in cardiovascular disease applications.

\section{Abstract}

Cardiac magnetic resonance (CMR) imaging, offering high-resolution images in a non-invasive format, serves as the golden standard for cardiac disease diagnosis and function assessment~\cite{wang2024screening}. Many machine learning approaches have emerged in medical imaging analysis, enabling effective information extraction and supporting various healthcare applications, such as disease detection and segmentation~\cite{isensee2018automatic, sander2020automatic}. However, many of these methods rely on annotations and labeled data, such as segmentation labels, which are time-consuming and labor-intensive to obtain. As a result, unsupervised learning methods have gained attention for their potential in medical image and organ function analysis~\cite{balakrishnan2018unsupervised,lv2021unsupervised} using only unlabeled data. However, the application of unsupervised learning to CMR imaging analysis, especially in cardiac function analysis and screening, remains underexplored. 

\begin{figure}
    \centering
    \includegraphics[width=0.9\linewidth]{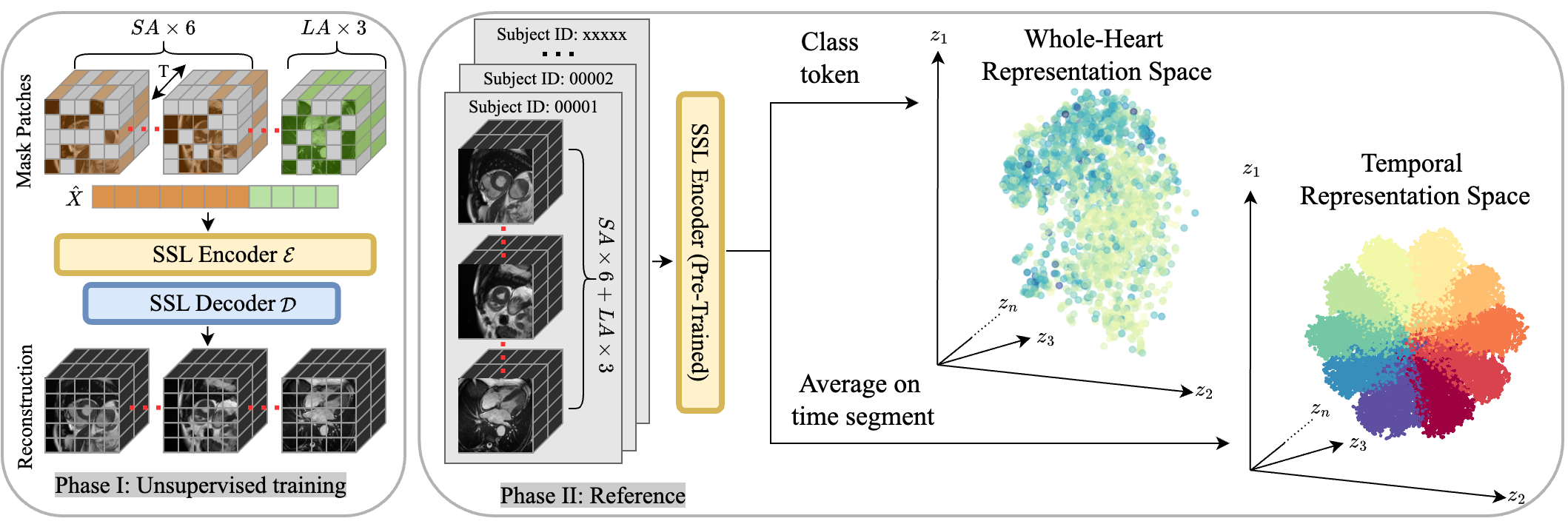}
    \caption{Overview of the proposed method. Phase I: Unsupervised learning is achieved through the reconstruction of a stack of multi-view masked 2D+T slices (6 SA and 3 LA). Phase II: We take the class token representing the whole-heart latent space and disentangled temporal latent representations by taking averages on different time segments.}
    \label{fig:archi}
\end{figure}

In this work, we present an unsupervised learning approach for high-level cardiac function assessment and screening. Our model generates well-structured and meaningful representations through random masking and disentangles temporal information to highlight the heart's motion characteristics. As shown in Fig~\ref{fig:archi}, our model~\cite{zhang2024whole} utilizes a masked autoencoder\cite{he2022masked} that processes multi-view 2D+T cardiac slices, where each slice is divided into small 3D blocks, with 70\% randomly masked. By reconstructing the full stack of slices, the model is encouraged to focus on high-level features of the entire heart and the cardiac cycle. Additionally, we disentangle 10 temporal embeddings from each subject, corresponding to different segments of the cardiac cycle. These time-specific embeddings emphasize the heart's motion and enable dynamic assessment of cardiac motion and function across various cycle phases.

The training process uses a dataset of 6,000 cardiac MR scans from the UK Biobank study\cite{petersen2015uk}. Each scan has 6 short-axis and 3 long-axis (2CH, 3CH, and 4CH) views 2D+T slices, each with a spatial resolution of (128, 128) and 50 time frames. We evaluate our model on an additional set of 6,000 scans from the same dataset. For visualization, we use t-SNE~\cite{maaten2008visualizing} with the number of embeddings set to 3 and a perplexity parameter of 5.

\begin{figure}
    \centering
    \includegraphics[width=0.9\linewidth]{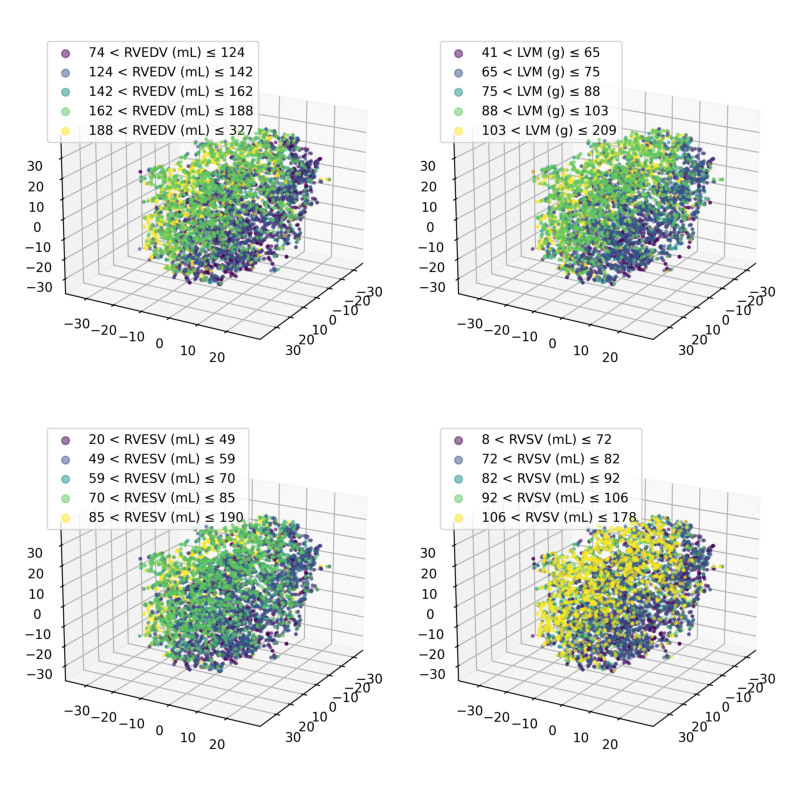}
    \caption{The 3D t-SNE visualization of the latent whole-heart representation obtained through reconstruction. Latent embeddings are labeled with right ventricular end-diastolic volume (RVEDV, top left), left ventricular mass (LVM, top right), right ventricular end-systolic volume (RVESV, bottom left), and right ventricular stroke volume (RVSV, bottom right) values, categorized into 5 groups according to the ground truth, and shown in different colors.}
    \label{fig:phenotypes}
\end{figure}

We first present a 3D t-SNE visualization of the learned latent representations for the entire heart, labeled according to key phenotype values—right ventricular end-diastolic volume (RVEDV), left ventricular mass (LVM), right ventricular end-systolic volume (RVESV), and right ventricular stroke volume (RVSV). These phenotypes are derived from the 3D spatial and temporal information of the heart. As shown in Fig.~\ref{fig:phenotypes}, our model produces a well-clustered latent space that effectively captures spatiotemporal variations among subjects.

\begin{figure}
    \centering
    \includegraphics[width=0.8\linewidth]{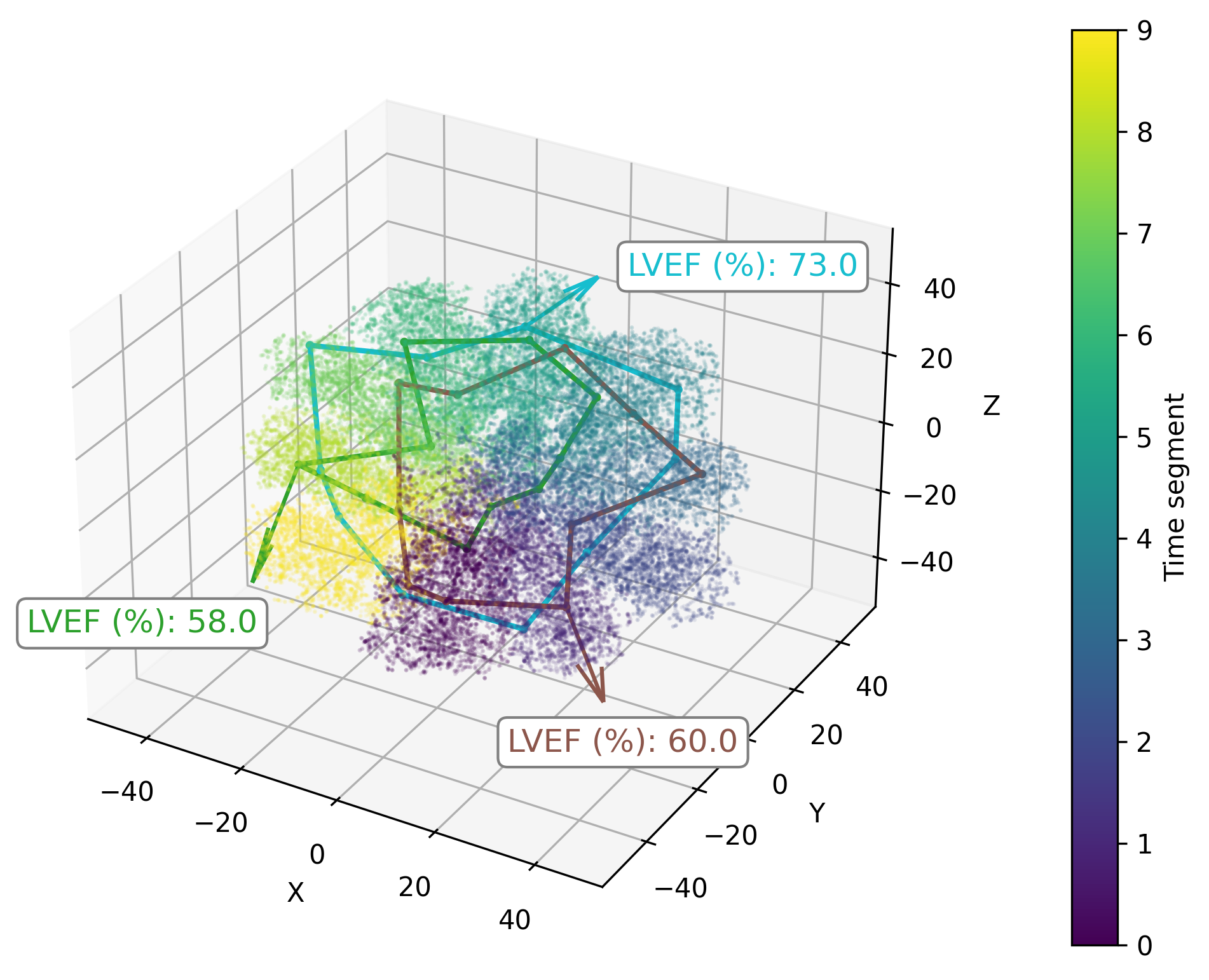}
    \caption{Temporal latent representations of 6,000 subjects with subject-specific trajectories. Each subject is represented by 10 distinct, color-coded temporal embeddings corresponding to specific time segments. Trajectories for three subjects are illustrated by connecting their 10 temporal embeddings, with the left ventricular ejection fraction (LVEF) values indicated.}
    \label{fig:trajectory}
\end{figure}

We further demonstrate the results of temporal representation disentanglement. In Fig.~\ref{fig:trajectory}, we visualize the disentangled temporal latent space, with each subject represented by 10 distinct, color-coded their corresponding to specific time segments and the quantitative results describing the clustering ability are shown in Tab.~\ref{tab:knn}. Clear clustering confirms that, despite substantial inter-subject anatomical variability at the pixel level, the model reliably captures temporal distinctions within each subject. Additionally, trajectories are formed by connecting each subject's temporal representations, with the left ventricular ejection fraction (LVEF) values indicated. As shown in Fig.~\ref{fig:trajectory}, we observe the subject with higher LVEF demonstrating larger trajectory perimeters. This suggests a link between phenotype values and trajectory characteristics, underscoring the model's potential for preliminary high-level cardiac function assessment and screening.

This work demonstrates the huge potential of preliminary cardiac screening and function assessment using unlabeled MR images. By generating a meaningful spatio-temporal latent space and further disentangled temporal latent space, our approach enables more targeted studies of time-specific cardiac function and shows great potential in diverse applications, such as classification and anomaly detection.
\begin{table}[ht!]
\centering
\setlength{\tabcolsep}{2mm}{}
\caption{The average percentage of neighboring temporal latent representations across all 6000 subjects that belong to the same temporal segment. k is the number of neighboring points that are checked out.}
\label{tab:knn}
\begin{tabular}{c|c|c|c|c}
\toprule
\#Neighbours & k=10 & k=100 & k=1000 & k=5000 \\ \hline
Mean over time & $100.0\%$ & $99.98\%$ & $99.11\%$ & $81.90\%$ \tstrut\\
\bottomrule
\end{tabular}
\end{table}

\newpage
\bibliographystyle{splncs04}
\bibliography{ref_ismrm2025.bib}

\end{document}